\documentclass[12pt]{article}

\usepackage{amsmath}
\usepackage{amssymb}
\usepackage{amsthm}
\usepackage{mathrsfs}
\usepackage[T1]{fontenc}
\usepackage{enumerate}
\usepackage{color}
\usepackage{cite}

\def\cond{condition}
\def\comrel{comutation relation}
\def\tfn{transformation}

\def\dd{Drinfel'd double}
\def\vbe{beta function equations}
\def\4diml{four-dimensional}
\def\bkg{background}
\def\wrt{with respect to}
\def\-1{^{-1}}
\def\half{\frac{1}{2}}
\def\coor{coordinate}

\def\cd{{\mathfrak d}}
\def\cg{{\mathfrak g}}
\def\tcg{\tilde{\mathfrak g}}
\def\hcg{\hat{\mathfrak g}}
\def\bcg{\bar{\mathfrak g}}

\def\wt{\tilde}
\def\wh{\widehat}
\def\wwt{\widetilde}

\def\pltp{Poisson--Lie T-pluralit}
\def\pltd{Poisson--Lie T-dualit}

\def\sugra{Generalized Supergravity Equation}
\def\dft{Double Field Theory}

\def\cf{{\mathcal {F}}}

\newcommand{\unit}{\mathbf{1}}
\newcommand{\nul}{\mathbf{0}}
\newcommand{\A}{\mathscr{A}}
\newcommand{\B}{\mathscr{B}}
\newcommand{\M}{\mathscr{M}}
\newcommand{\D}{\mathscr{D}}
\newcommand{\G}{\mathscr{G}}
\newcommand{\tG}{\widetilde{\mathscr{G}}}
\newcommand{\hG}{\widehat{\mathscr{G}}}
\newcommand{\bG}{\bar{\mathscr{G}}}

\newcommand{\J}{\mathcal{J}}

\newcommand{\hJ}{\widehat{\mathcal{J}}}
\newcommand{\bJ}{\bar{\mathcal{J}}}

\newcommand{\N}{\mathscr{N}}

\newcommand{\PL}{Poisson--Lie}

\newcommand{\wb}{\bar}

\newcommand{\cB}{\mathcal B}
\newcommand{\cD}{\mathcal D}
\newcommand{\cF}{\mathcal F}
\newcommand{\cH}{\mathcal H}

\title{T-folds as Poisson--Lie plurals}
\author{Ladislav Hlavat\'y\footnote{hlavaty@fjfi.cvut.cz}
\\ {\em Faculty of Nuclear Sciences and Physical Engineering,}
\\ {\em Czech Technical University in Prague,}
\\ {\em Czech Republic}
\and
Ivo Petr\footnote{ivo.petr@fit.cvut.cz}
\\ {\em Faculty of Information Technology,}
\\ {\em Czech Technical University in Prague,}
\\ {\em Czech Republic}}

\begin{document}
\maketitle

\begin{abstract}
In previous papers we have presented many purely bosonic solutions
of Generalized Supergravity Equations obtained by Poisson--Lie
T-duality and plurality of flat and Bianchi cosmologies. In this
paper we focus on their compactifications and identify solutions
that can be interpreted as T-folds. To recognize T-folds we adopt
the language of Double Field Theory and discuss how Poisson--Lie
T-duality/plurality fits into this framework. {As a special case we
confirm that all non-Abelian T-duals can be compactified as T-folds.}
\end{abstract}


\tableofcontents


\section{Introduction}

Dualities in string theory relate apparently different physical
models and allow to address issues that are otherwise hard to
tackle. T-duality \cite{buscher:ssbfe} connects models in \bkg s
with distinct curvature properties. Together with its non-Abelian
generalization \cite{delaossa:1992vc} it was extended to RR fields
\cite{1012.1320 sfethomp,1104.5196LCST} and can be used as solution
generating technique in supergravity \cite{KlebWitt,INNSZ,BTZiran,Goedeliran} and
generalized supergravity \cite{sugra2,Wulff:2016tju,Araujo:2017jkb}.
It also often contributes to the study of integrable models
\cite{BorWulff1,hoatsey}. However, most of the papers deal with
local aspects of non-Abelian duals, and global properties remain
unclear, see e.g. \cite{SomeGlobal}. The same holds for \pltd y
\cite{klise,klim:proc} or plurality \cite{unge:pltp} that introduce
\dd\ as the underlying algebraic structure of T-duality allowing us
to treat both original and dual/plural models equally.

Recently several papers \cite{sakamoto:yb,fernandez:Tfolds} appeared
that describe compactifications of Yang--Baxter deformations of
Minkowski and $\text{AdS}_5\times \textrm{S}^5$ \bkg s in terms of
T-folds. T-folds represent a special class of nongeometric
backgrounds that appeared in string theory in an attempt to
accommodate T-duality as symmetry of some models
\cite{Hellerman:2002ax,Dabholkar:2002sy,Kachru:2002sk,Flournoy:2004vn,hasslust,Lowe:2003qy}.
T-folds generalize the notion of manifold by allowing not only
diffeomorphisms but also T-duality transformations as transition
functions between local charts. Natural language for their
description has become Double Field Theory
\cite{Siegel:1993xq,Hull:2004in,Hull:2009mi,Hohm:2010pp}.
It turns out that Double Field Theory can describe not only Abelian T-duality but also \PL\ T-duality \cite{hassler,dehato,LuOst,saka2} and may help investigate quantum aspects of \PL\ T-duality \cite{borwulff:qplt} or its extension to U-duality \cite{sakatani:Udual,malekthom:pltU}.

In \cite{hlape:pltdid,hlape:pltpbia} we have presented many purely
bosonic solutions of \sugra s that were obtained by \pltd y or
plurality \tfn\ of flat and Bianchi cosmologies
\cite{Batakis,hokico}. The purpose of this paper is to present
solutions that can be interpreted as T-folds. We follow the idea
that T-folds can be identified using non-commutative structure
$\Theta$ in the open string picture \cite{seibergWitten}. We give
the argument both in terms of \PL\ T-plurality and \dft\ to show the
interplay between these two formalisms.  {From the structure of \dd\
underlying non-Abelian T-duality one finds that all non-Abelian
T-duals can be compactified as T-folds (as noticed e.g. in \cite{Bugden}). In the case of general plurality
\tfn\ additional
conditions have to be satisfied.}

{The paper is organized as follows. In Sections \ref{sec_pltp} and
\ref{SUGRA} we briefly recapitulate \pltp y and \sugra s.} Elements
of \dft, T-folds, and the method that we use to identify T-folds are
explained in Section \ref{sec_dft}. In Section \ref{secB5} we
present \bkg s obtained as \PL\ plurals of flat \bkg\ that can be
interpreted as T-folds. Examples of T-folds obtained as \PL\ plurals
of curved Bianchi cosmologies are presented in Section \ref{secB6}.


\section{Preliminaries}\label{sec_prel}

In this Section we will summarize main features of \pltd y and
plurality \cite{klise,klim:proc,unge:pltp,hlapevoj}, \sugra s
\cite{sugra2,Wulff:2016tju,Araujo:2017jkb} and T-folds
\cite{Dabholkar:2002sy,Hull:2004in,Hull:2009mi,Hohm:2010pp,fernandez:Tfolds,Plauschinn}.
For detailed information see the original papers.

\subsection{\pltd y/plurality}\label{sec_pltp}

A convenient way to describe \pltp y is in terms of a \dd. As this
has been done in many preceeding papers{,e.g. \cite{hlape:pltpbia}},
we shall not go into details and restrict to a summary of necessary
formulas.

Let $\G$ be a $d$-dimensional Lie group with free action on manifold
$\M$ of dimension {$M=n+d$. Since the action of $\G$ is transitive
on its orbits, we may locally consider $\M\approx (\M/\G) \times \G
= \N \times \G$. {This allows us to introduce the so-called} adapted
coordinates\footnote{{For a thorough description of the process of
finding adapted coordinates see} \cite{hlafilp:uniq}.}
\begin{equation}\nonumber\label{adapted}
\{s_\alpha,x^a\},\qquad \alpha=1, \ldots,n = \dim \N ,\ \ a=1,
\ldots, d = \dim \G
\end{equation}
{where $x^a$ denote group coordinates while $s_\alpha$ label the
orbits of $\G$ and will be treated as spectators in the
duality/plurality transformation.}

We shall consider sigma models on $\N \times \G$ given by
{covariant} tensor field $\cF$ invariant \wrt\ the action of group
$\G$. Such $\cF$ is defined by spectator-dependent $(n+d)\times
(n+d)$ matrix $E(s)$ and group dependent $\mathcal{E}(x)$ as
\begin{equation}\label{F}
\cf(s,x)=\mathcal{E}(x)\cdot E(s)\cdot \mathcal{E}^T(x), \qquad
\mathcal{E}(x)=
\left(
\begin{array}{cc}
 \unit_n & 0 \\
 0 & e(x)
\end{array}
\right).
\end{equation}
{The $d\times d$ matrix $e(x)$ contains} components of
right-invariant Maurer--Cartan form $(dg)g^{-1}$ on $\G$. The
dynamics of sigma model on $\M$ follows from Lagrangian
\begin{equation}\label{Lagrangian}
{\cal L}=\partial_- \phi^{\mu}\cf_{\mu\nu}(\phi)\partial_+
\phi^\nu,\qquad \phi^\mu=\phi^\mu(\sigma_+,\sigma_-), \qquad
\mu=1,\ldots,M=n+d
\end{equation}
where tensor field $\cf=\mathcal G + \mathcal B$ on $\M$ contains metric $\mathcal G$ and torsion potential (Kalb--Ramond field) $\mathcal B$.

To find \PL\ dual model \cite{klise} on $\N\times\tG$ we embed group
$\G$ into \dd, i.e. $2d$-dimensional Lie group $\D= (\G|\tG)$ formed
by a pair of Lie subgroups $\G$ and $\tG$. {The Lie algebra $\cd$ of
the \dd\ is endowed} with an ad-invariant non-degenerate symmetric
bilinear form $\langle . , . \rangle$. { $\cd$ can be written as a}
double cross sum  $\cg \bowtie \tcg$ of subalgebras $\cg$ and $\tcg$
\cite{majid} {corresponding to $\G$ and $\tG$. $\cg$ and $\tcg$} are
maximally isotropic with respect to {the form} $\langle . , .
\rangle$. {The resulting algebraic structure $(\cd, \cg, \tcg)$ is
called Manin triple.} {As noted already in \cite{klise}, for a
particular \dd\ $\D=(\G|\tG)$ there may exist various Manin
triples.} If {there is another pair of subgroups $\hG$ and $\bG$
(with corresponding Manin triple $(\cd,\hcg,\bcg)$) that form the
same \dd\ $\D$}, we can find \PL\ plural models \cite{unge:pltp} on
$\N \times \hG$ or $\N \times \bG$. {The \PL\ T-plural sigma model
on $\N \times \hG$ is specified} by tensor field
\begin{equation} \label{Fhat}
\widehat{\cf}(s,\hat x)= \mathcal{\widehat E}(\hat x)\cdot \widehat
E(s,\hat x) \cdot \mathcal{\widehat E}^T(\hat x), \qquad
\mathcal{\widehat E}(\hat x)=
\begin{pmatrix}
\unit_n & 0 \\
 0 & \wh e(\hat x)
\end{pmatrix}
\end{equation}
where $\wh e(\hat x)$ {contains components of $(d\hat g)\hat
g^{-1}$} on $\wh\G$,
\begin{equation}\label{Fhat2}
\wh E(s,\hat x)=\left(\unit_{n+d}+\wh E(s) \cdot \wh{\Pi}(\hat x)\right)^{-1}\cdot
\wh E(s) =\left(\wh E\-1(s)+ \wh{\Pi}(\hat x)\right)^{-1},
\end{equation}
$$
\widehat\Pi(\hat x)= \left(
\begin{array}{cc}
\nul_n & 0 \\
 0 & \widehat b(\hat x) \cdot \widehat a^{-1}(\hat x)
\end{array}
\right),
$$ and\footnote{The second identity in \eqref{Fhat2} holds for invertible $\wh E(s)$} {$\widehat b(\hat x)$, $\widehat a(\hat x)$ denote} submatrices of the adjoint representation of $\wh\G$ on {algebra $\mathfrak d$ given by}
$$
ad_{{\hat g}\-1}(\wb T) = \widehat b(\hat x) \cdot \wh T + \widehat a^{-1}(\hat x)\cdot\wb T.
$$
The matrix $\wh E(s)$ is obtained as follows. Let $C$ be an invertible $2d\times 2d$ matrix relating bases of Manin triples $\cg \bowtie \tcg$ and $\wh{\mathfrak g}\bowtie \wb{\mathfrak g}$ as
\begin{equation}\label{C_mat}
\begin{pmatrix}
\wh T \\
\wb T
\end{pmatrix}
 = C \cdot
\begin{pmatrix}
T \\
\widetilde T
\end{pmatrix} \qquad T_a \in \cg,\ \widetilde{T}^a \in \tcg,\ \wh T_a \in \hcg,\ \wb{T}^a \in
\bcg,\ a=1, \ldots, d.
\end{equation}
{We denote the $d \times d$ blocks of $C^{-1}$ as $P, Q, R, S$,
i.e.}
\begin{equation}\label{pqrs}
\begin{pmatrix}
T \\
\widetilde T
\end{pmatrix}
= C^{-1} \cdot
\begin{pmatrix}
\wh T \\
\wb T
\end{pmatrix} =
\begin{pmatrix}
 P & Q \\
 R & S
\end{pmatrix} \cdot
\begin{pmatrix}
\wh T \\
\wb T
\end{pmatrix}.
\end{equation}
{To account for the spectator fields we form} $(n+d)\times (n+d)$
matrices
\begin{equation}
\label{pqrs2}
\mathcal{P} =\begin{pmatrix}\unit_n &0 \\ 0&P \end{pmatrix}, \quad \mathcal{Q} =\begin{pmatrix}\nul_n&0 \\ 0&Q \end{pmatrix}, \quad \mathcal{R} =\begin{pmatrix}\nul_n&0 \\ 0&R \end{pmatrix}, \quad \mathcal{S} =\begin{pmatrix}\unit_n &0 \\ 0& S \end{pmatrix}.
\end{equation}
$\wh E(s)$ {then reads}
\begin{equation}\label{E0hat}
\wh E(s)=(\mathcal{P}+ E(s) \cdot \mathcal{R})^{-1} \cdot
(\mathcal{Q}+E(s) \cdot \mathcal{S}).
\end{equation}
This procedure was used in \cite{hlape:pltdid,hlape:pltpbia} to construct new solutions of \sugra s from Bianchi cosmologies.

Matrices $C$ relating Manin triples $\cg \bowtie \tcg$ and $\wh{\mathfrak g}\bowtie \wb{\mathfrak g}$ are not unique and different choices may lead to \bkg s with different curvature or torsion properties \cite{hlape:pltprev}.
However, in \cite{hlape:pltdid,hlape:pltpbia} we have shown that many parameters appearing in general $C$ are irrelevant as the resulting \bkg s differ only by a coordinate or gauge \tfn. {In such case we choose a representative of the class of $C$ matrices leading to such ``equivalent'' backgrounds.}

In this paper we deal with particular $C$'s leading to \bkg s that can be interpreted as T-folds. Important special cases include $P=S=\nul_d,\ Q=R=\unit_d$ in which case plurality reduces to \PL\ T-duality. For $P=S=\unit_d, \ R=\nul_d,\ Q=B$ where $B= - B^T$ we obtain the so-called $\cB$-shifts, while the so-called $\beta$-shifts are given by $P=S=\mathbf{1}_d,\ Q=\mathbf{0}_d$ and $R=\beta, \ \beta=-\beta^T$.

{Throughout this paper we deal with non-semisimple Bianchi groups
$\G$,} while $\tG$ is three-dimensional Abelian group $\A$. {We
parametrize group elements as} $g=e^{x^1 T_1}e^{x^2 T_2}e^{x^3 T_3}$
where $e^{x^2 T_2}e^{x^3 T_3}$ and $e^{x^3 T_3}$ parametrize normal
subgroups of $\G$}.  {\PL\ T-plurality transformation between models
on groups $\G$, $\tG$ and $\hG$, $\bG$ is specified by mapping
\eqref{C_mat}, but the relation can be also formulated in terms of
group elements as}
\begin{equation}
\label{ghgh}
l=g(y)\wt h(\wt y)=\wh g(\hat x)\bar h(\bar x),\quad l\in\D, \
g\in \G,\ \wt h\in \wt\G,\ \wh g\in \wh\G,\ \bar h\in \bar\G.
\end{equation}
{In this paper we consider Bianchi cosmologies on four-dimensional
manifolds, thus $\dim \N=1$. For simplicity we denote the spectator
coordinate $s_1$ by $t$.}

\subsection{Generalized Supergravity Equations}\label{SUGRA}

Adopting the convention used in \cite{hokico} we write the
Generalized Supergravity Equations of Motion
\cite{sugra2,Wulff:2016tju} as\footnote{We restrict to
{bosonic fields in the NS-NS sector.}}
\begin{align}\label{betaG}
0 &= R_{\mu\nu}-\frac{1}{4}H_{\mu\rho\sigma}H_{\nu}^{\
\rho\sigma}+\nabla_{\mu}X_{\nu}+\nabla_{\nu}X_{\mu},\\ \label{betaB}
0 &=
-\frac{1}{2}\nabla^{\rho}H_{\rho\mu\nu}+X^{\rho}H_{\rho\mu\nu}+\nabla_{\mu}X_{\nu}-\nabla_{\nu}X_{\mu},\\
\label{betaPhi} 0 &=
R-\frac{1}{12}H_{\rho\sigma\tau}H^{\rho\sigma\tau}+4\nabla_{\mu}X^{\mu}-4X_{\mu}X^{\mu}.
\end{align}
$H_{\rho\mu\nu}$ are components of torsion
$$
H_{\rho\mu\nu}=\partial_\rho \mathcal B_{\mu\nu}+\partial_\mu \mathcal B_{\nu\rho}+\partial_\nu \mathcal B_{\rho\mu},
$$
$\nabla_\mu$ are covariant derivatives \wrt\ metric $\mathcal G$, and $X$ is given by
\begin{equation} \label{Xform}
X_{\mu}=\partial_{\mu}\Phi + \J^{\nu} \cf_{\nu\mu},
\end{equation}
where $\Phi$ is the dilaton {and $ \J$ is Killing vector of the \bkg\ $\cf$ \cite{Wulff:2016tju}, i.e.
$$
\mathcal{L}_{\J}\cf=0.
$$}
When vector field $ \J$ vanishes the usual \vbe\ are recovered.

{Transformation formulas for $\Phi$ and $\J$ in the context of \pltp y were given in \cite{unge:pltp,dehato,saka2,hlape:pltpbia}.
In this paper we shall use these formulas in the form  that take into account
possible non-locality of the resulting dilaton $\wh\Phi$. 
We set
\begin{equation}\label{phi0}
\Phi^{0}(y)=\Phi(y)+\half \ln\Big| \det \left[\left({\bf 1} + \Pi(y)
E(s)\right) a(y)\right]\Big|,
\end{equation}
and express $y$ (\coor s of $\G$) from \eqref{ghgh} in terms of $\hat x$ and $\bar x$. If the dependence is linear
\begin{equation}
\nonumber
 y^k= \hat d_m^k \hat x^m+ \bar d^{k m}\bar x_m,
\end{equation}
the transformed dilaton reads
\begin{equation}
\wh\Phi(\hat x)= \Phi^{0}(\hat d_m^k \hat x^m)-\half\ln \Big| \det \left[\left(N +
\wh\Pi(\hat x) M\right)\wh a(\hat x)\right]\Big|
\label{dualdil}
\end{equation}
where
$$ M=\mathcal{S}^T E(s)-\mathcal{Q}^T\,, \qquad
N=\mathcal{P}^T-\mathcal{R}^T E(s).
$$
Components of vectors $\wh{\mathcal{J}}$ for \bkg s on $\N\times\wh\G$ are
$$
\wh{\mathcal{J}}^\alpha=0,\qquad \alpha=1,\ldots, \dim \N,
$$
\begin{equation}\label{kilJ}
\mathcal{\wh J}^{\dim \N+ m} = \left(\half{{\bar f}^{ab}}{}_b -\frac{\partial{\Phi^{0}}}{\partial y^k}\bar d^{k a}\right){\wh V_a}^m, \qquad a,b,k,m=  1,
\ldots,\dim \G,
\end{equation}
where $\wh V_a$ are left-invariant fields on $\hG$ and ${\bar f}^{ab}{}_c$ are structure constants of $\bG$.}

\subsection{Short review of Double Field Theory and T-folds}\label{sec_dft}

The presence of dualities in string theory suggests that it might be convenient to generalize the standard concept of manifold.
T-folds do so by allowing T-dualities beside diffeomorphisms as transition functions between charts.
To describe T-folds, we shall use the framework of \dft\ (DFT).

While T-folds are examples of the so-called nongeometric backgrounds \cite{Plauschinn}, in DFT T-duality transformations become diffeomorphisms of a manifold with doubled dimension.
In this construction all the local patches are geometric.

The central concepts in DFT are the generalized metric $\cH\in\mathcal{O}(M,M)$ and DFT dilaton $\cD$ that are defined using the initial \bkg\ fields $\mathcal G, \mathcal B, \Phi$ as
\begin{equation}\label{genmet}
\cH  =
\begin{pmatrix}
\mathcal G - \mathcal B\cdot\mathcal G\-1\cdot\mathcal B & \mathcal B\cdot\mathcal G\-1 \\
-\mathcal G\-1\cdot\mathcal B & \mathcal G\-1
\end{pmatrix}
\end{equation}
and
\begin{equation}
\label{dftdilaton}
\mathcal D=\Phi-\frac{1}{4}\ln(\det\mathcal G).
\end{equation}

The possibility to compactify both initial and extended
$2M$-dimensional manifold in the direction of {the vector field $\alpha$} is \cond ed by
existence of a monodromy matrix {$\Omega$} whose action is
equivalent to {action of the vector field $\alpha$
\begin{equation}\label{Hmonodromy2}
(e^{\cal L_\alpha}\rhd\cH)(x^\mu)=\Omega\rhd\cH(x^\mu):=\Omega  \cdot \cH(x^\mu) \cdot {\Omega }^T, 
\end{equation}
and invariance of DFT dilaton
\begin{equation}
\label{dftshift2} (e^{\cal L_\alpha}\rhd\mathcal D)(x^\mu)= \mathcal D(x^\mu),
\end{equation}
where $\cal L_\alpha$ is the Lie derivative in the direction of $\alpha$.
In the adapted \coor s for $\alpha$, which we assume here, this action is 
a shift of \coor s in direction $\alpha^{\mu}$, i.e.
\begin{equation}\label{Hmonodromy}
 \cH(x^\mu+\alpha^\mu)=\Omega  \cdot \cH (x^\mu) \cdot {\Omega }^T,
\end{equation}
\begin{equation}
\label{dftshift} \mathcal D(x^\mu+\alpha^\mu)= \mathcal D(x^\mu)
\end{equation}
where $\Omega$ is a constant matrix.}

If the \tfn \ by $\Omega$ can be reinterpreted as \tfn \ of
the \bkg \ $\cf= \mathcal G + \mathcal B$ induced by \coor \ or
gauge \tfn\ ($\cB$-shift), then $\cf$ can be considered as \bkg \ on compactified
manifold $\M$. Such cases will be considered trivial in the rest of the paper and we shall not discuss them further. However, monodromy matrix of the form
\begin{equation}\label{Ibeta}
\Omega_{\beta}= \left(
\begin{array}{cc}
\unit_M & 0 \\
\beta & \unit_M
\end{array}
\right), \qquad
\beta^T=-\beta
\end{equation}
where $\beta$ is constant antisymmetric $M\times M$ matrix cannot be
obtained neither by \tfn\ of \coor s on $\M$ nor gauge \tfn\ of
$\mathcal B$ and tensor $\cf$ must be interpreted as \bkg\ on a
T-fold.

Authors of paper \cite{fernandez:Tfolds} give simple procedure for
T-fold identification via open\footnote{called dual in
\cite{fernandez:Tfolds}} \bkg\ \cite{seibergWitten}. To understand
the action of $\Omega_{\beta}$ we rewrite the generalized metric
\eqref{genmet} in terms of open fields $G$ and $\Theta$ as
\begin{equation}\label{genmetopen}
\cH  =
\begin{pmatrix}
G & - G \cdot\Theta \\
\Theta\cdot G & G\-1 - \Theta\cdot G \cdot \Theta
\end{pmatrix}.
\end{equation}
{where $\Theta $ is the antisymmetric part of $\cf^{-1}(s,x)$,
i.e.
\begin{equation}
\nonumber\label{bivectorTheta}
\Theta^{\mu\nu}=-((\mathcal G + \mathcal B)\-1\cdot \mathcal B\cdot(\mathcal G -\mathcal B)\-1)^{\mu\nu}=-\Theta^{\nu\mu},
\end{equation}
and 
$$
G_{\mu\nu}=(\mathcal G - \mathcal B\cdot\mathcal G\-1\cdot\mathcal B)_{\mu\nu}=G_{\nu\mu}.
$$
These two tensors form the so-called open \bkg\ \cite{seibergWitten}
and the form of bivector $\Theta$ is important for identification of
T-folds.}

It is easy to verify that action of $\Omega_{\beta}$ given by \eqref{Ibeta}  does not
change $G$, but shifts $\Theta$ by $\beta$. Necessary \cond\ \eqref{Hmonodromy} then {reads}
\begin{equation}\label{ThetaG_cond}
\Theta(x^\mu+\alpha^\mu)=\Theta(x^\mu)+\beta, \qquad
G(x^\mu+\alpha^\mu)=G(x^\mu).
\end{equation}
It can be satisfied only for {\bkg s $\cf$ where $\Theta$ are} linear in a \coor\ $x^\mu$ and
suitable matrices $\beta$ can be obtained from $\Theta$ as linear
combinations of the so-called Q-fluxes \cite{LuOst}
\begin{equation}\label{mtxbeta}
 \beta^{\mu\nu}=
 \alpha^\lambda\,\partial_\lambda\Theta^{\mu\nu}
 =\alpha^\lambda\,{Q_\lambda}^{\mu\nu}.
\end{equation}
Here, the sum \eqref{mtxbeta} runs only over $\lambda$ for which
the open metric $G$ is invariant \wrt\ $x^\lambda\rightarrow
x^\lambda +\alpha^\lambda$. {Conclusion then is that \bkg s with constant nonvanishing Q-fluxes can be globally defined as T-folds.}

In terms of the \bkg\ on $\M$ the condition \eqref{Hmonodromy} is
then equivalent to
\begin{equation}\label{betashift}
 \cf\-1(x^\mu+\alpha^\mu)=\cf\-1(x^\mu)+\beta.
\end{equation}
The right-hand side can be considered as {\PL\ \tfn\ \eqref{E0hat} by $ \beta$-shift}. {For quantized strings one needs $\Omega_{\beta}\in \mathcal O (M,M, \mathbb{Z})$, which is satisfied if entries of the antisymmetric matrix $\beta$ are integers.}

In \cite{hlape:pltdid,hlape:pltpbia} we investigated \PL\ duals and
plurals of Bianchi cosmologies. To find \bkg s that can be
interpreted as T-folds we  will present in the rest of this paper
dual or plural \bkg s whose bivector $\Theta$ is linear in \coor s
$x^\mu$. As the plural \bkg s are given by \eqref{Fhat},
\eqref{Fhat2}, \eqref{E0hat}, bivector $\wh \Theta$ can be expressed
as
\begin{equation}
\label{hatTheta}
\wh\Theta(\wh x)=\mathcal{\widehat V}(\hat x) \cdot
\left(\half\left(\wh E\-1(s)-\wh E^{-T}(s)\right)+ \wh{\Pi}(\hat x)\right) \cdot
\mathcal{\widehat V}(\hat x)^T,
\end{equation}
$$
\mathcal{\widehat V}(\hat x)=
\left(
\begin{array}{cc}
 \unit_n & 0 \\
 0 & \widehat v(\hat x)
\end{array}
\right)
$$
where $\widehat v(\hat x)$ is $d\times d$ matrix of components of
right-invariant vector fields of the group $\widehat\G$. From the
formula \eqref{hatTheta} it is clear that bivectors $\wh\Theta$
linear in $\hat x$ occur beside others for backgrounds on Abelian
groups $\wh\G=\A$. For such groups $\mathcal{\widehat V}(\hat
x)=\unit_M$ and $\wh{\Pi}$ is linear in $\hat x$ since
\begin{equation} \label{Pi=fx}\wh\Pi(\wh x)=\left(
\begin{array}{cc} \mathbf 0_n & 0 \\ 0 & \wh b(\wh x) \end{array}\right),\quad \wh b_{ab}(\wh x)={\bar f_{ab}}^c \, \wh x_c.
\end{equation}
Moreover, the open metric $\wh G$ is completely independent of the
group coordinates $\hat x$. It means that {all \bkg s obtained
by non-Abelian T-duality}, i.e. on a semi-Abelian \dd\ $\D=(\G|\A)$,
{can be compactified as T-folds in coordinates $\hat x$.} In a different way this has
been shown also in \cite{Bugden}.


\section{Transformations of flat metric}\label{secB5}

{In this section we} investigate  T-folds obtained from \pltp ies of
the Minkowski metric following from its invariance \wrt\ Bianchi
groups \cite{PWZ}.

\subsection{T-folds obtained by \PL\ \tfn s given by Bianchi $V$ isometry}\label{secBV}

The flat Minkowski metric is invariant \wrt\ the action of Bianchi
$V$ group. {In adapted \coor s it has the form}\footnote{{To get
matrix $E(s)$ one has to set $y_1=0$ in $\cf$}.}
\begin{equation} \label{F5_orig}
\cf (t,y_1) = \left(
\begin{array}{cccc}
 -1 & 0 & 0 & 0 \\
 0 & t^2 & 0 & 0 \\
 0 & 0 & e^{2 y_1} t^2 & 0 \\
 0 & 0 & 0 & e^{2 y_1} t^2 \\
\end{array}
\right).
\end{equation}
Together with vanishing dilaton $\Phi = 0$ the background satisfies
\vbe, i.e. equations \eqref{betaG}--\eqref{betaPhi} with $\J = 0$.
Its non-Abelian dual \wrt\ {non-semisimple} Bianchi $V$ group
appears repeatedly in the literature as it is not conformal
\cite{GRV, EGRSV} but satisfies \sugra s
{\cite{fernandez:Tfolds}}.
The flat background in the form
\eqref{F5_orig} can be obtained  {from \eqref{F} by virtue of}
six-dimensional semi-Abelian \dd\footnote{{We use ${\B}_{V}$ and
${\mathfrak b}_{V}$ to} denote the group Bianchi ${V}$ and its Lie
algebra. {Other Bianchi groups and algebras are denoted similarly.
The three-dimensional Abelian group and its Lie algebra are written
as $\A$ and $\mathfrak a$.}} $\D=({\B}_V |\A)$ {with corresponding
Manin triple} $\mathfrak{d}=\mathfrak b_V\bowtie \mathfrak a$.

\subsubsection{Transformation of $\mathfrak b_{V}\bowtie \mathfrak a$ to $\mathfrak b_{VI_{-1}}\bowtie \mathfrak b_{Vii}$}

The algebra $\mathfrak{d}=\mathfrak b_V\bowtie \mathfrak a$ allows
several other decompositions into Manin triples
\cite{snohla:ddoubles}, such as $\mathfrak{d}= \mathfrak
b_{VI_{-1}}\bowtie \mathfrak b_{Vii}$. Interested reader may find
its commutation relations in \cite{hlape:pltpbia}.
{Two matrices $C$ transforming Manin triple $\mathfrak
b_{V}\bowtie \mathfrak a$ to $\mathfrak b_{VI_{-1}}\bowtie \mathfrak
b_{Vii}$ and producing geometrically different \bkg s are}
\begin{equation} \nonumber 
C_1=\left(
\begin{array}{cccccc}
 -1 & 0 & 0 & 0 & 0 & 0 \\
 0 & 1 & 0 & 0 & 0 & 0 \\
 0 & 0 & 0 & 0 & 0 & 1 \\
 0 & 1 & 0 & -1 & 0 & 0 \\
 1 & 0 & 0 & 0 & 1 & 0 \\
 0 & 0 & 1 & 0 & 0 & 0 \\
\end{array}
\right), \qquad C_2=\left(
\begin{array}{cccccc}
 1 & 0 & 0 & 0 & 0 & 0 \\
 0 & 0 & 0 & 0 & 1 & 0 \\
 0 & 0 & 1 & 0 & 0 & 0 \\
 0 & 0 & 0 & 1 & 1 & 0 \\
 -1 & 1 & 0 & 0 & 0 & 0 \\
 0 & 0 & 0 & 0 & 0 & 1 \\
\end{array}
\right).
\end{equation}

{Using $C_1$ we get background tensor}
\begin{equation}\label{mtz51615ii1}
 \wh\cf(t,\hat x_1,\hat x_3) = \left(
\begin{array}{cccc}
 -1 & 0 & 0 & 0 \\
 0 & \frac{e^{2 \hat x_1} \left(\hat x_3^2+1\right) t^2}{t^4+e^{2 \hat x_1} \left(\hat x_3^2+1\right)} & \frac{t^4}{t^4+e^{2 \hat x_1} \left(\hat x_3^2+1\right)} & \frac{e^{2 \hat x_1} \hat x_3 t^2}{t^4+e^{2 \hat x_1} \left(\hat x_3^2+1\right)} \\
 0 & -\frac{t^4}{t^4+e^{2 \hat x_1} \left(\hat x_3^2+1\right)} & \frac{t^2}{t^4+e^{2 \hat x_1} \left(\hat x_3^2+1\right)} & \frac{e^{2 \hat x_1} \hat x_3}{t^4+e^{2 \hat x_1} \left(\hat x_3^2+1\right)} \\
 0 & \frac{e^{2 \hat x_1} \hat x_3 t^2}{t^4+e^{2 \hat x_1} \left(\hat x_3^2+1\right)} & -\frac{e^{2 \hat x_1} \hat x_3}{t^4+e^{2 \hat x_1} \left(\hat x_3^2+1\right)} & \frac{e^{2 \hat x_1} \left(t^4+e^{2 \hat x_1}\right)}{t^2 \left(t^4+e^{2 \hat x_1} \left(\hat x_3^2+1\right)\right)} \\
\end{array}
\right)
\end{equation}
whose metric is curved and torsion does not vanish. {Supported by
dilaton}
$$
\wh\Phi(t,\hat x_1,\hat x_3)=-\frac{1}{2} \ln \left(t^6 e^{-4\hat x_1}+t^2 e^{-2 {\hat x_1}}\left(\hat x_3^2+1\right)\right)
$$
the \bkg\ satisfies \sugra s with {constant vector} $\hJ
=(0,0,2,0)$.

DFT metric then has the form \small
$$ \left(
\begin{array}{cccccccc}
 -1 & 0 & 0 & 0 & 0 & 0 & 0 & 0 \\
 0 & t^2 & 0 & 0 & 0 & 0 & t^2 & 0 \\
 0 & 0 & e^{-2 \hat x_1} t^2 & 0 & 0 & -e^{-2 \hat x_1} t^2 & 0 &
 e^{-2 \hat x_1} t^2 \hat x_3 \\
 0 & 0 & 0 & \frac{e^{2 \hat x_1}}{t^2} & 0 & 0 & -\frac{e^{2
 \hat x_1} \hat x_3}{t^2} & 0 \\
 0 & 0 & 0 & 0 & -1 & 0 & 0 & 0 \\
 0 & 0 & -e^{-2 \hat x_1} t^2 & 0 & 0 & e^{-2 \hat x_1}
 t^2+\frac{1}{t^2} & 0 & -e^{-2 \hat x_1} t^2 \hat x_3 \\
 0 & t^2 & 0 & -\frac{e^{2 \hat x_1} \hat x_3}{t^2} & 0 & 0 &
 t^2+\frac{e^{2 \hat x_1} \left(\hat x_3^2+1\right)}{t^2} & 0 \\
 0 & 0 & e^{-2 \hat x_1} t^2 \hat x_3 & 0 & 0 & -e^{-2 \hat x_1} t^2
 \hat x_3 & 0 & e^{-2 \hat x_1} t^2 \left(\hat x_3^2+1\right) \\
\end{array}
\right)$$\normalsize
and DFT dilaton is
$$ \mathcal D (t,\hat x_1)= -\frac{1}{4}\ln(t^6)+ \hat x_1.$$
Bivector $\wh\Theta$ for the \bkg\ above is of the form
\begin{equation}
\label{theta51615ii}
 \wh\Theta=\left(
\begin{array}{cccc}
 0 & 0 & 0 & 0 \\
 0 & 0 & -1 & 0 \\
 0 & 1 & 0 & -{\hat x_3} \\
 0 & 0 & {\hat x_3} & 0 \\
\end{array}
\right)\end{equation}
so that the Q-flux is constant. Its nonvanishing components are
\begin{equation}\label{qflux51615ii2}
{Q_4}^{34}=-{Q_4}^{43}=-1
\end{equation}
and $\beta$-shift matrix is
\begin{equation}\label{betashift51615ii2}
\beta=\left(
\begin{array}{cccc}
 0 & 0 & 0 & 0 \\
 0 & 0 &0 & 0 \\
 0 & 0 & 0 & -\alpha \\
 0 & 0 & \alpha & 0 \\
\end{array}
\right).
\end{equation}
$\beta$-shift \eqref{betashift} of $\wh\cf$ is then equivalent to
the \coor\ shift $\hat x_3\mapsto \hat x_3+\alpha$. DFT dilaton  is
independent of $\hat x_3$ so the \cond\ \eqref{dftshift}
is satisfied. Therefore, the \bkg\
\eqref{mtz51615ii1} can be compactified as a $T$-fold as
$$(t,\hat x_1, \hat x_2, \hat x_3)\sim(t, \hat x_1,\hat x_2, \hat x_3+\alpha).$$
Note that none of the relevant fields depend on $\hat x_2$ and it is possible to compactify also in $\hat x_2$ to obtain a $T^2$-fold. Such possibilities are disregarded in the rest of the paper to emphasize nontrivial compactifications.

{Using $C_2$ we obtain plural \bkg\ that reads}
\begin{equation}\label{mtz51615ii2}
 \wh\cf(t,\hat x_1,\hat x_3) =\left(
\begin{array}{cccc}
 -1 & 0 & 0 & 0 \\
 0 & \frac{e^{2 \hat x_1} \left(\hat x_3^2+1\right) t^2}{e^{2 \hat x_1} \left(\hat x_3^2+1\right)+1} & \frac{1}{e^{2 \hat x_1} \left(\hat x_3^2+1\right)+1} & \frac{e^{2 \hat x_1} \hat x_3 t^2}{e^{2 \hat x_1} \left(\hat x_3^2+1\right)+1} \\
 0 & -\frac{1}{e^{2 \hat x_1} \left(\hat x_3^2+1\right)+1} & \frac{1}{\left(e^{2 \hat x_1} \left(\hat x_3^2+1\right)+1\right) t^2} & \frac{e^{2 \hat x_1} \hat x_3}{e^{2 \hat x_1} \left(\hat x_3^2+1\right)+1} \\
 0 & \frac{e^{2 \hat x_1} \hat x_3 t^2}{e^{2 \hat x_1} \left(\hat x_3^2+1\right)+1} & -\frac{e^{2 \hat x_1} \hat x_3}{e^{2 \hat x_1} \left(\hat x_3^2+1\right)+1} & \frac{e^{2 \hat x_1} \left(1+e^{2 \hat x_1}\right) t^2}{e^{2 \hat x_1} \left(\hat x_3^2+1\right)+1} \\
\end{array}
\right).
\end{equation}
Its metric is curved but torsion vanishes. {The \vbe\ are satisfied
for dilaton}
$$
\wh\Phi(t,\hat x_1,\hat x_3)=-\frac{1}{2} \ln \left(t^2 \left(1+e^{2 {\hat x_1}} \left(\hat x_3^2+1\right)\right)\right).
$$

Bivector $\wh\Theta$ is the same as in the preceding case and DFT dilaton is $\mathcal D (t,\hat x_1)= -\frac{1}{4}\ln(t^6) - \hat x_1$. The \bkg\ \eqref{mtz51615ii2} can be again compactified as $T$-fold by
$$(t, \hat x_1, \hat x_2, \hat x_3)\sim(t, \hat x_1, \hat x_2, \hat x_3+\alpha).$$

\subsubsection{Transformation of $\mathfrak b_{V}\bowtie \mathfrak a$ to $\mathfrak b_{II}\bowtie \mathfrak b_{V}$}

Examples of mappings {producing geometrically different \bkg s} that transform the algebraic structure of Manin
triple $\mathfrak b_{V}\bowtie \mathfrak a$ to $\mathfrak
b_{II}\bowtie \mathfrak b_{V}$ are
\begin{equation} \nonumber C_1=\left(
\begin{array}{cccccc}
 0 & 0 & 0 & 1 & 0 & 0 \\
 0 & 0 & \frac{1}{2} & 0 & 1 & 0 \\
 0 & -\frac{1}{2} & 0 & 0 & 0 & 1 \\
 1 & 0 & 0 & 0 & 0 & 0 \\
 0 & 1 & 0 & 0 & 0 & 0 \\
 0 & 0 & 1 & 0 & 0 & 0 \\
\end{array}
\right), \qquad C_2=\left(
\begin{array}{cccccc}
 0 & 0 & 0 & -1 & 0 & 0 \\
 0 & 1 & 0 & 0 & 0 & \frac{1}{2} \\
 0 & 0 & 1 & 0 & -\frac{1}{2} & 0 \\
 -1 & 0 & 0 & 0 & 0 & 0 \\
 0 & 0 & 0 & 0 & 1 & 0 \\
 0 & 0 & 0 & 0 & 0 & 1 \\
\end{array}
\right).
\end{equation}
Background obtained by $C_1$ is given by tensor
\begin{align}\label{mtz51251}
& \wh\cf(t,\hat x_2,\hat x_3)= \\ & \nonumber \left(
\begin{array}{cccc}
 -1 & 0 & 0 & 0 \\
 0 & \frac{t^2}{t^4+\hat x_2^2+\hat x_3^2} & \frac{\hat x_3 t^2+2 \hat x_2}{2 \left(t^4+\hat x_2^2+\hat x_3^2\right)} & \frac{\hat x_2 t^2+2\hat  x_3}{2 \left(t^4+\hat x_2^2+\hat x_3^2\right)} \\
 0 & \frac{t^2 \hat x_3-2 \hat x_2}{2 \left(t^4+\hat x_2^2+\hat x_3^2\right)} & \frac{\left(\hat x_3^2+4\right) t^4+4 \hat x_3^2}{4 t^2
   \left(t^4+\hat x_2^2+\hat x_3^2\right)} & \frac{-2 t^6+\hat x_2 \hat x_3 t^4-4 \hat x_2^2 t^2-4 \hat x_2 \hat x_3}{4 t^2 \left(t^4+\hat x_2^2+\hat x_3^2\right)} \\
 0 & \frac{t^2 \hat x_2-2 \hat x_3}{2 \left(t^4+\hat x_2^2+\hat x_3^2\right)} & \frac{2 t^6+\hat x_2 \hat x_3 t^4+4 \hat x_2^2 t^2-4 \hat x_2 \hat x_3}{4 t^2
   \left(t^4+\hat x_2^2+\hat x_3^2\right)} & \frac{\left(\hat x_2^2+4\right) t^4+4 \hat x_2^2}{4 t^2 \left(t^4+\hat x_2^2+\hat x_3^2\right)} \\
\end{array}
\right)
\end{align}
with curved metric and nonvanishing torsion. Together with {
$$
\wh\Phi(t,\hat x_2,\hat x_3)=-\frac{1}{2} \ln \left(t^2 \left(t^4+\hat x_2^2+\hat x_3^2\right)\right), \qquad \hJ =(0,2,0,0)
$$
the \bkg\ satisfies \sugra s.}

Bivector $\wh\Theta$ has the form
$$\wh\Theta=\left(
\begin{array}{cccc}
 0 & 0 & 0 & 0 \\
 0 & 0 & \frac{\left(t^4-4\right) \hat x_2}{t^4+4} & -\hat x_3 \\
 0 & -\frac{\left(t^4-4\right) \hat x_2}{t^4+4} & 0 & \frac{2 t^4}{t^4+4} \\
 0 & \hat x_3 & -\frac{2 t^4}{t^4+4} & 0 \\
\end{array}
\right)$$ and is clearly linear in both $\hat x_2$ and $\hat x_3$.
However, components of Q-flux corresponding to shift in $\hat x_2$
depend on $t$ and the open metric $\wh G$ depends on $\hat x_2$ so
the above given procedure does not apply. Constant nonvanishing
components of Q-flux are
\begin{equation}\label{qflux51251}
{Q_4}^{24}=-{Q_4}^{42}=-1,
\end{equation}
and $\beta$-shift \eqref{betashift} of $\wh\cf$ given by
\begin{equation}\label{betashift51251}
\beta=\left(
\begin{array}{cccc}
 0 & 0 & 0 & 0 \\
 0 & 0 &0 & -\alpha \\
 0 & 0 & 0 &0 \\
 0 & \alpha & 0 & 0 \\
\end{array}
\right)
\end{equation}
is equivalent to the \coor\ shift $\hat x_3 \mapsto \hat x_3+\alpha$. DFT dilaton
\begin{equation}
\label{dftdil51251}
\mathcal D(t)=-\frac{3}{4} \ln \left(t^2\right)
\end{equation}
is independent of $\hat x_3$. Therefore, the \bkg\ \eqref{mtz51251} can be
compactified as $T$-fold by $$(t,\hat x_1,\hat x_2,\hat x_3)\sim(t,\hat x_1,\hat x_2,\hat x_3+\alpha). $$

Background obtained by $C_2$ is given by tensor
\begin{align}\label{mtz51251b}
& \wh\cf(t,\hat x_2,\hat x_3)= \\ & \nonumber \left(
\begin{array}{cccc}
 -1 & 0 & 0 & 0 \\
 0 & \frac{1}{t^2 \left(\hat x_2^2+\hat x_3^2+1\right)} & \frac{2 \hat x_2 t^2+\hat x_3}{2 t^2 \left(\hat x_2^2+\hat x_3^2+1\right)} & \frac{2 \hat x_3 t^2+\hat x_2}{2 t^2
   \left(\hat x_2^2+\hat x_3^2+1\right)} \\
 0 & \frac{\hat x_3-2 t^2 \hat x_2}{2 t^2 \left(\hat x_2^2+\hat x_3^2+1\right)} & \frac{4 \left(\hat x_3^2+1\right) t^4+\hat x_3^2}{4 t^2
   \left(\hat x_2^2+\hat x_3^2+1\right)} & \frac{-4 \hat x_2 x_3 t^4-2 \left(2 \hat x_2^2+1\right) t^2+\hat x_2 \hat x_3}{4 t^2 \left(\hat x_2^2+\hat x_3^2+1\right)} \\
 0 & \frac{\hat x_2-2 t^2 \hat x_3}{2 t^2 \left(\hat x_2^2+\hat x_3^2+1\right)} & \frac{-4 \hat x_2 \hat x_3 t^4+\left(4 \hat x_2^2+2\right) t^2+\hat x_2 \hat x_3}{4 t^2
   \left(\hat x_2^2+\hat x_3^2+1\right)} & \frac{4 \left(\hat x_2^2+1\right) t^4+\hat x_2^2}{4 t^2 \left(\hat x_2^2+\hat x_3^2+1\right)} \\
\end{array}
\right)
\end{align}
with curved metric and vanishing torsion.
Together with dilaton
$$
\wh\Phi(t,\hat x_2,\hat x_3)=-\frac{1}{2} \ln \left(t^2\left(\hat x_2^2+\hat x_3^2+1\right)\right)
$$
they satisfy \vbe.

Bivector $\wh\Theta$ for this \bkg \ is of the form
$$
\wh\Theta=\left(
\begin{array}{cccc}
 0 & 0 & 0 & 0 \\
 0 & 0 & \frac{\left(1-4 t^4\right) {\hat x_2}}{4 t^4+1} & -{\hat x_3}
 \\
 0 & \frac{\left(4 t^4-1\right) {\hat x_2}}{4 t^4+1} & 0 &
 \frac{2}{4 t^4+1} \\
 0 & {\hat x_3} & -\frac{2}{4 t^4+1} & 0 \\
\end{array}
\right).
$$
It is slightly different from the previous case, however, the constant nonvanishing components of Q-flux, $\beta$-shift matrix and DFT dilaton are the same as in \eqref{qflux51251}, \eqref{betashift51251} and \eqref{dftdil51251}, and the \bkg\ \eqref{mtz51251b} can be compactified as $T$-fold by
$$
(t,\hat x_1,\hat x_2,\hat x_3)\sim(t,\hat x_1,\hat x_2,\hat x_3+\alpha).
$$

\subsection{T-folds obtained by \PL\ \tfn s given by Bianchi $IV$ isometry}

Next we investigate \bkg s that follow from \pltp ies obtained from the invariance of Minkowski metric \wrt\ the action of Bianchi $IV$ group. The group ${\B}_{IV}$ is not semisimple and trace of
its structure constants does not vanish. The metric in adapted \coor s reads
\begin{equation} \label{F4_orig}
\cf (t,y_1) = \left(
\begin{array}{cccc}
 1 & 0 & 0 & 0 \\
 0 & 0 & e^{-y_1} y_1 & e^{-y_1} \\
 0 & e^{-y_1} y_1 & e^{-2 y_1} & 0 \\
 0 & e^{-y_1} & 0 & 0 \\
\end{array}
\right).
\end{equation}
Since the \bkg\ is flat and torsionless, the \vbe\ are satisfied if we choose vanishing dilaton $\Phi = 0$.
In this form the \bkg\ can be obtained by formula \eqref{F} if we consider
six-dimensional semi-Abelian \dd\ $\D=({\B}_{IV} |\A)$ and Manin triple
$\mathfrak{d}=\mathfrak b_{IV}\bowtie \mathfrak a$ spanned by basis
$(T_1,T_2,T_3,\wwt T^1,\wwt T^2,\wwt T^3)$ with non-trivial \comrel s
\begin{align}
\label{MT41}
[ T_1, T_2] & = -T_2+T_3, & [ T_1, T_3] & = -T_3, & [T_1, \wwt T^2] & = \wwt T^2, \\ \nonumber
[ T_1, \wwt T^3] & = -\wwt T^2+\wwt T^3, & [T_2, \wwt T^2] & = -\wwt T^1, & [T_2, \wwt T^3] & = \wwt T^1, & [ T_3, \wwt T^3] & = -\wwt T^1.
\end{align}
The algebra $\cd$ allows several other decompositions into Manin triples.

\subsubsection{Transformation of $\mathfrak b_{IV}\bowtie \mathfrak a$ to $\mathfrak b_{VI_{-1}}\bowtie \mathfrak b_{II}$}

Lie algebra $\mathfrak{d}= \mathfrak b_{VI_{-1}}\bowtie \mathfrak b_{II}$ is spanned by basis $(\wh T_1,\wh T_2,\wh T_3,\wb T^1,\wb
T^2,\wb T^3)$ with algebraic relations
\begin{align}
\nonumber [ \wh T_1, \wh T_2] & =-\wh T_2, & [ \wh T_1, \wh T_3] & = \wh T_3,\\
[ \wh T_1, \wb T^2] & = \wh T_3+\wb T^2, & [ \wh T_1, \wb T^3] & = -\wh T_2 -\wb T^3, & [ \wh T_2, \wb T^2] & = -\wb T^1, \\
\nonumber [ \wh T_3, \wb T^3] & = \wb T_1, & [ \wb T^2, \wb T^3] & =
\wb T^1.
\end{align}
Mapping that transforms the algebraic structure of
Manin triple $\mathfrak b_{IV}\bowtie \mathfrak a$ to $\mathfrak b_{VI_{-1}}\bowtie \mathfrak b_{II}$ is given by matrix
\begin{equation} \label{C416-12}
 C_1=\left(
\begin{array}{cccccc}
 -1 & 0 & 0 & 0 & 0 & 0 \\
 0 & 0 & 0 & 0 & 1 & 0 \\
 0 & 0 & -1 & 0 & 0 & 0 \\
 0 & 0 & 0 & -1 & 0 & 0 \\
 0 & 1 & 0 & 0 & 0 & 0 \\
 0 & 0 & 0 & 0 & 0 & -1 \\
\end{array}
\right).
\end{equation}
Resulting background tensor
reads
\begin{equation}\label{mtz416121}
\wh\cf(t, \hat x_1) = \left(
\begin{array}{cccc}
 1 & 0 & 0 & 0 \\
 0 & -\hat x_1^2 & -e^{-\hat x_1} \hat x_1 & e^{\hat x_1}
 \\
 0 & e^{-\hat x_1} \hat x_1 & e^{-2 \hat x_1} & 0 \\
 0 & e^{\hat x_1} & 0 & 0 \\
\end{array}
\right).
\end{equation}
The \bkg\ has vanishing torsion. Its metric is curved with vanishing scalar curvature. Simple coordinate transformation brings the metric to the Brinkmann form of a plane-parallel wave
\begin{equation}
ds^2=\frac{2 z_3^2}{u^2}du^2+2du\,dv+d z_3^2+d z_4^2.
\end{equation}
Together with the dilaton
$$
\wh\Phi(\hat x_1)=-\hat x_1,
$$
the \bkg\ satisfies \vbe.

Bivector $\wh\Theta$ for this \bkg\  is of the form
$$
\wh\Theta=\left(
\begin{array}{cccc}
 0 & 0 & 0 & 0 \\
 0 & 0 & 0 & 0 \\
 0 & 0 & 0 & -\hat x_1 \\
 0 & 0 & \hat x_1 & 0 \\
\end{array}
\right).$$ Nonvanishing components of constant Q-flux are
\begin{equation}\label{qflux416122}
{Q_2}^{34}=-{Q_2}^{43}=-1,
\end{equation}
so the $\beta$-shift matrix is
\begin{equation}\label{betashift416122}
\beta=\left(
\begin{array}{cccc}
 0 & 0 & 0 & 0 \\
 0 & 0 &0 & 0 \\
 0 & 0 & 0 &-\alpha \\
 0 & 0 &\alpha & 0 \\
\end{array}
\right).
\end{equation}
Unfortunately, DFT dilaton $\mathcal D$ and the open metric depend
on $\hat x_1$, {i.e. the \cond\ \eqref{dftshift} and second part of \eqref{ThetaG_cond} are not satisfied. Therefore, the \bkg\ \eqref{mtz416121} cannot be
compactified as $T$-fold.
This is also true for \bkg\ obtained by the other matrix
$$
C_2=\left(
\begin{array}{cccccc}
 1 & 0 & 0 & 0 & 0 & 0 \\
 0 & 0 & -1 & 0 & 0 & 0 \\
 0 & 0 & 0 & 0 & 1 & 0 \\
 0 & 0 & 0 & 1 & 0 & 0 \\
 0 & 0 & 0 & 0 & 0 & -1 \\
 0 & 1 & 0 & 0 & 0 & 0 \\
\end{array}
\right)
$$
that transforms $\mathfrak b_{IV}\bowtie \mathfrak a$ to $\mathfrak b_{VI_{-1}}\bowtie \mathfrak b_{II}$ since it only gives the \bkg\ \eqref{mtz416121} in different \coor s.} A bit different situation holds in the following case. 

\subsubsection{Transformation of $\mathfrak b_{IV}\bowtie \mathfrak a$ to $\mathfrak b_{II}\bowtie\mathfrak b_{VI_{-1}}$}

Manin triple $\mathfrak b_{II}\bowtie\mathfrak b_{VI_{-1}}$ is dual
to that of preceding section. Background given by matrix
\begin{equation}\nonumber
C_1=\left(
\begin{array}{cccccc}
 0 & 0 & 0 & -1 & 0 & 0 \\
 0 & 1 & 0 & 0 & 0 & 0 \\
 0 & 0 & 0 & 0 & 0 & -1 \\
 -1 & 0 & 0 & 0 & 0 & 0 \\
 0 & 0 & 0 & 0 & 1 & 0 \\
 0 & 0 & -1 & 0 & 0 & 0 \\
\end{array}
\right)
\end{equation}
that is dual to \eqref{C416-12} has the form
\begin{equation}\label{mtz416121d}
 \wb\cf(t, \bar x_2,\bar x_3) = \left(
\begin{array}{cccc}
1 & 0 & 0 & 0 \\
 0 & 0 & 0 & \frac{1}{\bar x_3+1} \\
 0 & 0 & 1 & \frac{\bar x_2}{\bar x_3+1} \\
 0 & \frac{1}{1-\bar x_3} & \frac{\bar x_2}{\bar x_3-1} & \frac{(\bar x_2-2) \bar x_2}{\bar x_3^2-1} \\
\end{array}
\right).
\end{equation}
{Formulas \eqref{dualdil}, \eqref{kilJ}} for transformation of dilaton and supplementary Killing vector $\bJ$ give
\begin{equation}
\label{dilandJ}
\wb\Phi(t, \bar x_3)=-\frac{1}{2} \ln \left(-1+\bar x_3^2\right)
\qquad \bJ =(0,1,0,0).
\end{equation}
Together with this nontrivial dilaton and $\bJ$ the \bkg\ satisfies
Generalized Supergravity Equations. The \bkg\ has flat metric and
vanishing torsion so it also satisfies \vbe\ with vanishing dilaton.
This is rather interesting situation. Dilaton $\bar\Phi$ and $\bJ$
enter \sugra s combined into one-form $X$ and there might be more
suitable combinations.

Bivector $\bar \Theta$ is of the form
$$
\bar\Theta=\left(
\begin{array}{cccc}
 0 & 0 & 0 & 0 \\
 0 & 0 & \bar x_2 & -\bar x_3 \\
 0 & -\bar x_2 & 0 & 0 \\
 0 & \bar x_3 & 0 & 0 \\
\end{array}
\right)
$$
and DFT dilaton $\mathcal D$ vanishes.
Nonvanishing components of Q-flux are
\begin{equation}
{Q_3}^{23}=-{Q_3}^{32}=1,\quad {Q_4}^{24}=-{Q_4}^{42}=-1
\end{equation}
and it may seem that $\beta$-shift matrices can be arbitrary
linear combinations
\begin{equation}
\beta=\alpha_2 \left(
\begin{array}{cccc}
 0 & 0 & 0 & 0 \\
 0 & 0 & 1 & 0 \\
 0 & -1 & 0 & 0 \\
 0 & 0 & 0 & 0 \\
\end{array}
\right)+\alpha_3 \left(
\begin{array}{cccc}
 0 & 0 & 0 & 0 \\
 0 & 0 & 0 & -1 \\
 0 & 0 & 0 & 0 \\
 0 & 1 & 0 & 0 \\
\end{array}
\right).
\end{equation}
However, the open metric $\bar G$ depends linearly on $\bar x_2$.
Therefore, \bkg\ \eqref{mtz416121d} can be compactified as $T$-fold
{only by}
$$
(t,\bar x_1, \bar x_2, \bar x_3)\sim(t, \bar x_1, \bar x_2, \bar x_3+\alpha_3).
$$
{This also  happens for the other matrix
$$ C_2=\left(
\begin{array}{cccccc}
 0 & 0 & 0 & 1 & 0 & 0 \\
 0 & 0 & 0 & 0 & 0 & -1 \\
 0 & 1 & 0 & 0 & 0 & 0 \\
 1 & 0 & 0 & 0 & 0 & 0 \\
 0 & 0 & -1 & 0 & 0 & 0 \\
 0 & 0 & 0 & 0 & 1 & 0 \\
\end{array}
\right)
$$
that transforms $\mathfrak b_{IV}\bowtie \mathfrak a$ to $\mathfrak b_{II}\bowtie \mathfrak b_{VI_{-1}}$ giving \bkg\ \eqref{mtz416121d} in different coordinates.}

\section{Transformations of curved cosmologies}\label{secB6}

\subsection{T-folds obtained by transformation of Bianchi $VI_{-1}$ cosmology}\label{B_VI-1}

Let us now focus on the curved \bkg\ with metric \cite{hokico}
\begin{equation}
\label{mtz61}
\cf(t,y_1)=\left(
\begin{array}{cccc}
 -e^{-4 \Phi(t)}a_1(t)^2{a_2}(t)^4 & 0 & 0 & 0 \\
 0 &a_1(t)^2 & 0 & 0 \\
 0 & 0 & e^{-2 y_1}{a_2}(t)^2 & 0 \\
 0 & 0 & 0 & e^{2 y_1}{a_2}(t)^2 \\
\end{array}
\right)
\end{equation}
{that is invariant \wrt\ the action of Bianchi $VI_{_{-1}}$ group.
Dilaton
$$
\Phi(t) = \beta t
$$
and functions $a_i(t)$ are given as in \cite{Batakis} by}
\begin{equation} \nonumber
a_1(t) = \sqrt{p_1} \exp \left(\frac{1}{2} e ^{2 p_2 t} + \frac{p_1 t}{2} +\Phi(t)\right), \qquad a_2(t) = \sqrt{p_2} e^{\frac{p_2 t}{2}+ \Phi(t)}.
\end{equation}
{The \vbe\ are satisfied if parameters $p_1, p_2$ and $\beta$
fulfill condition
$$
\beta^2=\frac{1}{4}(2 p_1 p_2 + p_2^2).
$$}

{To obtain metric \eqref{mtz61} via \eqref{F} we shall consider
Manin triple $\mathfrak{d}=\mathfrak b_{VI_{-1}}\bowtie \mathfrak
a$}
{that corresponds to the same \dd\ that we discussed in Section
\ref{secBV}.}
{Although the \dd\ is the same, the \bkg\ in this Section is
different as we use different matrix $E(t)=\cf(t,y_1=0)$.}

\subsubsection{Transformation of $\mathfrak b_{VI_{-1}}\bowtie \mathfrak a$ to $\mathfrak a\bowtie\mathfrak b_{V} $}

Example of mappings that transform the algebraic structure of Manin
triple $\mathfrak b_{VI_{-1}}\bowtie \mathfrak a$ to $\mathfrak
a\bowtie\mathfrak b_{V}$ reads
 \begin{equation}
C=\left(
\begin{array}{cccccc}
 0 & 0 & 0 & -1 & 0 & 0 \\
 0 & 0 & 1 & 0 & 0 & 0 \\
 0 & 0 & 0 & 0 & 1 & 0 \\
 -1 & 0 & 0 & 0 & 0 & 0 \\
 0 & 0 & 0 & 0 & 0 & 1 \\
 0 & 1 & 0 & 0 & 0 & 0 \\
\end{array}
\right)
 \end{equation}
and background obtained by \pltp y with this matrix is given by tensor
\begin{equation}\label{mtz61516}
\wh \cf(t,\hat x_2,\hat x_3) =
\left(
\begin{array}{cccc}
 -e^{-4 t \beta } a_1^2 a_2^4 & 0 & 0 & 0 \\
 0 & \frac{a_2^2}{\Delta} & \frac{\hat x_2 a_2^4}{\Delta} & \frac{\hat x_3}{\Delta} \\
 0 & -\frac{\hat x_2 a_2^4}{\Delta} & \frac{a_2^2 \left(a_1^2 a_2^2+\hat x_3^2\right)}{\Delta} & -\frac{\hat x_2 \hat x_3 a_2^2}{\Delta} \\
 0 & -\frac{\hat x_3}{\Delta} & -\frac{\hat x_2 \hat x_3 a_2^2}{\Delta} & \frac{a_1^2+\hat x_2^2 a_2^2}{\Delta} \\
\end{array}
\right)
\end{equation}
where
$$
\Delta=a_1(t)^2 a_2(t)^2+\hat x_2^2 a_2(t)^4+\hat x_3^2.
$$
The \bkg\ is curved and has nontrivial torsion. Together with dilaton
$$
\wh \Phi(t,\hat x_2,\hat x_3)=\beta t-\frac{1}{2} \ln \Delta
$$
they satisfy \sugra s with $\hJ =(0,1,0,0)$.

Bivector $\wh \Theta$ for the \bkg\ \eqref{mtz61516} is of the form
$$
\wh \Theta=\left(
\begin{array}{cccc}
 0 & 0 & 0 & 0 \\
 0 & 0 & -\hat x_2 & -\hat x_3 \\
 0 & \hat x_2 & 0 & 0 \\
 0 & \hat x_3 & 0 & 0 \\
\end{array}
\right)
$$
and the Q-flux is constant. Its nonvanishing components are
\begin{equation}\label{qflux61516}
{Q_3}^{23}=-{Q_3}^{32}=-1,\quad {Q_4}^{24}=-{Q_4}^{42}=-1
\end{equation}
and $\beta$-shift matrices are arbitrary linear combinations
\begin{equation}\label{betashift61516}
\beta=\alpha_2 \left(
\begin{array}{cccc}
 0 & 0 & 0 & 0 \\
 0 & 0 & -1 & 0 \\
 0 & 1 & 0 & 0 \\
 0 & 0 & 0 & 0 \\
\end{array}
\right)+\alpha_3 \left(
\begin{array}{cccc}
 0 & 0 & 0 & 0 \\
 0 & 0 & 0 & -1 \\
 0 & 0 & 0 & 0 \\
 0 & 1 & 0 & 0 \\
\end{array}
\right).
\end{equation}
DFT dilaton is
$$
\mathcal D(t) = -\ln (a_1(t)a_2(t)^2)+2 \beta t
.
$$
$\beta$-shift \eqref{betashift} of $\wh\cf$ is then equivalent to the \coor\ shift $(\hat x_2,\hat x_3)\mapsto(\hat x_2+\alpha_2 ,\hat x_3+\alpha_3)$. Therefore, the \bkg\ \eqref{mtz61516} can be compactified as $T^2$-fold by
$$
(t,\hat x_1,\hat x_2,\hat x_3)\sim(t,\hat x_1,\hat x_2+\alpha_2
,\hat x_3+\alpha_3).
$$
For $\alpha_2,\alpha_3\in \mathbb{Z}$ we have $\Omega_{\beta}\in \mathcal O(M,M,\mathbb{Z})$.

\subsection{T-folds obtained by transformation of Bianchi $VI_\kappa$ cosmology}\label{secB6k}

Finally we are going {discuss} \pltp ies of curved cosmology
invariant \wrt\ Bianchi $VI_\kappa$ group. Its metric is
\cite{hokico}
\begin{equation} \label{mtz6kappa1}
\cf(t,y_1)=\left(
\begin{array}{cccc}
 -e^{-4 \Phi(t)}a_1(t)^2{a_2}(t)^2{a_3}(t)^2 & 0 & 0 & 0 \\
 0 &a_1(t)^2 & 0 & 0 \\
 0 & 0 & e^{2\kappa y_1}{a_2}(t)^2 & 0 \\
 0 & 0 & 0 & e^{2 y_1}{a_3}(t)^2 \\
\end{array}
\right)
\end{equation}
and dilaton reads
$$
\Phi(t) = \beta t.
$$
The functions $a_i(t)$ {are given as in\cite{Batakis} by}
\begin{align}\label{B6k_ai}
{a_1}(t)&=e^{\Phi(t)}\left(\frac{{p_1}}{\kappa +1}\right)^{\frac{\kappa^2+1}{(\kappa +1)^2}}e^{\frac{(\kappa -1) {p_2} t}{2 (\kappa +1)}}\sinh ^{-\frac{\kappa ^2+1}{(\kappa+1)^2}}({p_1} t),\nonumber\\
{a_2}(t)&= e^{\Phi(t)}\left(\frac{{p_1}}{\kappa +1}\right)^{\frac{\kappa}{\kappa +1}} e^{\frac{{p_2} t}{2}} \sinh ^{-\frac{\kappa}{\kappa +1}}({p_1} t),\\
{a_3}(t)&= e^{\Phi(t)}\left(\frac{{p_1}}{\kappa
+1}\right)^{\frac{1}{\kappa +1}}e^{-\frac{{p_2} t}{2}} \sinh
^{-\frac{\kappa}{\kappa +1}}({p_1} t).\nonumber
\end{align}
{The \vbe\ are satisfied provided constants $p_1, p_2, \beta$ and
$\kappa$ fulfill condition}
\begin{equation}\label{betapcond}
\beta^2=\frac{ \left(\kappa^2+\kappa +1\right) p_1^2}{(\kappa
+1)^2}-\frac{p_2^2}{4}.
\end{equation}

\subsubsection{Transformation of $\mathfrak b_{VI_\kappa}\bowtie \mathfrak a$ to $\mathfrak b_{VI_\kappa}\bowtie \mathfrak b_{VI_{-\kappa}.iii}$}

{The \bkg\ \eqref{mtz6kappa1} can be obtained from \eqref{F} if we
consider six-dimensional semi-Abelian \dd\ $\D=(\mathscr
B_{VI_\kappa}|\mathscr A)$. Among the decompositions of
$\mathfrak{d}=\mathfrak b_{VI_\kappa}\bowtie \mathfrak a$ given in
\cite{snohla:ddoubles} is also Manin triple $\mathfrak{d}=\mathfrak
b_{VI_\kappa}\bowtie \mathfrak b_{VI_{-\kappa}.iii}$.}
For its commutation relations see \cite{hlape:pltpbia} where
general forms of mappings $C$ relating $\mathfrak
b_{VI_\kappa}\bowtie \mathfrak a$ and $\mathfrak
b_{VI_\kappa}\bowtie \mathfrak b_{VI_{-\kappa}.iii}$ were found as
well. In particular we shall consider matrices
\begin{equation} \nonumber 
C_1=\left(
\begin{array}{cccccc}
 -1 & 0 & 0 & 0 & 0 & 0 \\
 0 & 0 & 0 & 0 & 1 & 0 \\
 0 & 0 & 0 & 0 & 0 & 1 \\
 0 & 0 & 0 & -1 & 0 & 1 \\
 0 & 1 & 0 & 0 & 0 & 0 \\
 1 & 0 & 1 & 0 & 0 & 0 \\
\end{array}
\right), \qquad
C_2=\left(
\begin{array}{cccccc}
 1 & 0 & 0 & 0 & 0 & 0 \\
 0 & 1 & 0 & 0 & 0 & 0 \\
 0 & 0 & 1 & 0 & 0 & 0 \\
 0 & 0 & 1 & 1 & 0 & 0 \\
 0 & 0 & 0 & 0 & 1 & 0 \\
 -1 & 0 & 0 & 0 & 0 & 1 \\
\end{array}
\right).
\end{equation}
Using matrix $C_1$ we get {\bkg}
\begin{align} \label{mtz6k6kiiia}
& \wh{\cf}(t,\hat x_1,\hat x_2)=\\ \nonumber & \left(
\begin{array}{cccc}
 -e^{-4 \beta t} a_1^2 a_2^2 a_3^2 & 0 & 0 & 0 \\
 0 & \frac{a_1^2 \left(a_2^2 a_3^2+e^{2 (\kappa +1) \hat x_1} \kappa ^2 \hat x_2^2\right)}{\Delta} & \frac{e^{2 (\kappa +1) \hat x_1} \kappa a_1^2 \hat x_2}{\Delta} & \frac{e^{2 \hat x_1} a_1^2 a_2^2}{\Delta} \\
 0 & \frac{e^{2 (\kappa +1) \hat x_1} \kappa a_1^2 \hat x_2}{\Delta} & \frac{e^{2 \kappa \hat x_1} \left(e^{2 \hat x_1} a_1^2+a_3^2\right)}{\Delta} & -\frac{e^{2 (\kappa +1) \hat x_1} \kappa \hat x_2}{\Delta} \\
 0 & -\frac{e^{2 \hat x_1} a_1^2 a_2^2}{\Delta} & \frac{e^{2 (\kappa +1) \hat x_1} \kappa \hat x_2}{\Delta} & \frac{e^{2 \hat x_1} a_2^2}{\Delta} \\
\end{array}
\right)
\end{align}
where
$$\Delta = e^{2 \hat x_1} a_1^2 a_2^2+a_3^2 a_2^2+e^{2 (\kappa +1) \hat x_1} \kappa ^2 \hat x_2^2.
$$
{For constant vector $\wh{\mathcal J}=(0,0,0,\kappa)$ and dilaton
\begin{equation} \nonumber 
\wh\Phi(t,\hat x_1,\hat x_2)= \beta t-\frac{1}{2} \ln
\left(\Delta\right)+(\kappa +1) \hat x_1
\end{equation}
the Generalized Supergravity Equations are satisfied
provided condition \eqref{betapcond} holds.}

From the matrix $C_2$ we get
\begin{align} \label{mtz6k6kiiib}
& \wh{\cf}(t,\hat x_1,\hat x_2)=\\ \nonumber &
\left(
\begin{array}{cccc}
-e^{-4 t \beta } a_1^2 a_2^2 a_3^2 & 0 & 0 & 0 \\
 0 & \frac{a_1^2 \left(\hat x_2^2 \kappa ^2 a_2^2 a_3^2 e^{2 \hat x_1 (\kappa +1)}+1\right)}{\Delta} & \frac{\hat x_2 \kappa  a_1^2 a_2^2 a_3^2 e^{2 \hat x_1 (\kappa +1)}}{\Delta} & \frac{e^{2 \hat x_1} a_1^2 a_3^2}{\Delta} \\
 0 & \frac{\hat x_2 \kappa  a_1^2 a_2^2 a_3^2 e^{2 \hat x_1 (\kappa +1)}}{\Delta} & \frac{a_2^2 e^{2 \hat x_1 \kappa } \left(e^{2 \hat x_1} a_1^2 a_3^2+1\right)}{\Delta} & -\frac{\hat x_2 \kappa  a_2^2 a_3^2 e^{2 \hat x_1 (\kappa +1)}}{\Delta} \\
 0 & -\frac{e^{2 \hat x_1} a_1^2 a_3^2}{\Delta} & \frac{\hat x_2 \kappa  a_2^2 a_3^2 e^{2 \hat x_1 (\kappa +1)}}{\Delta} & \frac{e^{2 \hat x_1} a_3^2}{\Delta} \\
\end{array}
\right)
\end{align}
where
$$\Delta = e^{2 \hat x_1} a_1^2 a_3^2+\hat x_2^2 \kappa ^2 a_2^2 a_3^2 e^{2 \hat x_1 (\kappa +1)}+1.
$$
Generalized Supergravity Equations with $\wh{\mathcal J}=(0,0,0,-1)$ are satisfied by dilaton
\begin{equation} \nonumber 
\wh\Phi(t,\hat x_1,\hat x_2)= \beta t-\frac{1}{2} \ln \left(\Delta\right)
\end{equation}
under the condition
\eqref{betapcond}.

Bivector $\wh\Theta$ and matrix $\beta$ for both \bkg s read
$$
\wh\Theta=
\left(
\begin{array}{cccc}
 0 & 0 & 0 & 0 \\
 0 & 0 & 0 & -1 \\
 0 & 0 & 0 & \kappa\, \hat x_2 \\
 0 & 1 & -\kappa \, \hat x_2 & 0 \\
\end{array}
\right), \qquad
\beta=\left(
\begin{array}{cccc}
 0 & 0 & 0 & 0 \\
 0 & 0 & 0 &0 \\
 0 & 0 & 0 & \alpha\,\kappa \\
 0 & 0 & -\alpha\,\kappa & 0 \\
\end{array}
\right),
$$
and nonvanishing components of Q-flux are
\begin{equation}
{Q_3}^{34}=-{Q_3}^{43}=\kappa.
\end{equation}
Since the corresponding DFT dilatons
$$
\mathcal D(t, \hat x_1) =-\ln (a_1(t)a_2(t)a_3(t))+2 \beta t \pm \frac{(\kappa+1)\hat x_1}{2}
$$
do not depend on $\hat x_2$, $\beta$-shifts \eqref{betashift} of
$\wh\cf$ are equivalent  to \coor\ shift $\hat x_2\mapsto \hat
x_2+\alpha$ and \bkg s \eqref{mtz6k6kiiia}, \eqref{mtz6k6kiiib} can be
compactified as $T$-folds by
$$
(t,\hat x_1,\hat x_2,\hat x_3)\sim(t,\hat x_1,\hat x_2+\alpha,\hat x_3).
$$

\section{Conclusions}

In this paper we discussed the connection between \PL\ T-plurality
and \dft. Using tensors $\Theta$ and $G$ constituting the open \bkg\
{one obtains
conditions \eqref{ThetaG_cond} and \eqref{betashift}}
that need to be satisfied in order to identify \bkg\ as T-fold, {namely that 
$\Theta$ given by the formula
\eqref{hatTheta} must be linear in a coordinate $x^{\mu}$.  The shift
$x^{\mu}\mapsto x^{\mu}+\alpha^\mu$ then is equivalent to a
$\beta$-shift \eqref{betashift}} provided $G$ and DFT dilaton $ \cD$
are independent of $x^{\mu}$. {The formula
\eqref{hatTheta} also implies that \bkg s obtained by non-Abelian T-duality can be
compactified as T-folds.} In Sections \ref{secB5} and \ref{secB6} we
have tested \cond s \eqref{ThetaG_cond} and \eqref{betashift} for
\PL\ plurals of flat and Bianchi cosmologies obtained in
\cite{hlape:pltpbia}. We have shown that in spite of their rather
complicated forms, many \PL\ plurals can be considered as T-folds.


\end{document}